\documentclass{article}
\usepackage{dcase2021,amsmath,graphicx,url,times,booktabs, tabularx, subcaption, caption}
\usepackage{adjustbox}
\usepackage{amssymb}
\usepackage{kotex}

\usepackage{xcolor}         
\usepackage{multirow}
\newcommand\sh[1]{\textcolor{black} {#1}} 
\newcommand\sm[1]{\textcolor{black} {#1}} 

\newcommand\wh[1]{\textcolor{black} {#1}} 

\usepackage{tikz}
\def\checkmark{\tikz\fill[scale=0.4](0,.35) -- (.25,0) -- (1,.7) -- (.25,.15) -- cycle;}

\title{Domain Generalization on Efficient Acoustic Scene Classification using Residual Normalization}

\name{Byeonggeun Kim$^1$, Seunghan Yang$^1$, Jangho Kim$^{1,2,*}$, Simyung Chang$^1$}
\address{$^1$Qualcomm AI Research${}^{\dagger}$\thanks{  ${}^{\dagger}$ Qualcomm AI Research is an initiative of Qualcomm Technologies, Inc.${}^{*}$Author completed the research in part during an internship at Qualcomm Technologies, Inc.}, Qualcomm Korea YH, Seoul, Republic of Korea\\
$^2$Seoul National University, Seoul, Republic of Korea\\
      \{kbungkun, seunghan, jangkim, simychan\}@qti.qualcomm.com}
\begin{document}

\ninept
\maketitle

\begin{sloppy}

\begin{abstract}

It is a practical research topic how to deal with multi-device audio inputs by a single acoustic scene classification system with efficient design. In this work, we propose Residual Normalization, a novel feature normalization method that uses \wh{frequency-wise} normalization 
with a shortcut path to discard unnecessary device-specific information without losing useful information for classification. Moreover, we introduce an efficient architecture, BC-ResNet-ASC, a modified version of the baseline architecture with a limited receptive field. BC-ResNet-ASC outperforms the baseline architecture even though it contains the small number of parameters. Through three model compression schemes: pruning, quantization, and knowledge distillation, we can reduce model complexity further while mitigating the performance degradation. The proposed system achieves an average test accuracy of 76.3\% in TAU Urban Acoustic Scenes 2020 Mobile, development dataset with 315k parameters, and average test accuracy of 75.3\% after compression to 61.0KB of non-zero parameters. The proposed method won the 1st place in DCASE 2021 challenge, TASK1A.
\end{abstract}


\section{Introduction}
\label{sec:intro}

Acoustic scene classification (ASC) is the task of classifying sound scenes such as ``airport'', ``train station'', and ``urban park'' to which a user belongs. ASC is an important research field that plays a key role in various applications such as context-awareness and surveillance \cite{valenti2016dcase,radhakrishnan2005audio,chu2009environmental}. Detection and Classification of Acoustic Scenes and Events (DCASE) \cite{dcase2021web} is an annual challenge, attracting attention to the field. 
There are various interesting tasks in the DCASE2021 challenge, and we aim for \sh{TASK1A:} Low-Complexity Acoustic Scene Classification with Multiple Devices \cite{dcase_task1A, dcase_dataset}.

TASK1A classifies \sm{ten} different audio scenes from 12 European cities using \sm{four} real and 11 simulated devices. In this year, the task becomes more challenging as an ASC model needs to solve two problems simultaneously which practically exist in real applications; First, data is collected from multiple devices\sm{,} and the number of samples per device is unbalanced.
Therefore, the proposed system needs to solve the domain imbalance problem while \sh{generalizing to different devices.}
Second, TASK1A restricts the model size and therefore requires an efficient network design.

In recent years, a number of researches have been proposed for more efficient and high-performance ASC.
\sh{Most of them are based on convolutional neural network (CNN)}
using residual network and ensemble \cite{task1a2020best_cnn, receptivefield, task1a2020_2nd_cnn, task1a2019best}.
The top-performing models in the previous TASK1A utilize multiple CNNs in a single model with parallel connections \cite{task1a2020best_cnn, task1a2020_2nd_cnn}.
For the generalization of the model, \cite{receptivefield, phaye2019subspectralnet} show that there is a regularization effect by adjusting the receptive field size in CNN-based design.
However, these works also use models of several MB, and it is still challenging to satisfy the low model complexity of TASK1A of this year. In addition, when using the previous methods, we found an accuracy drop of up to 20\% on the unseen devices compared to the device with sufficient training data.
In this work, we propose methods to leverage the generalization capabilities of unseen devices while maintaining the model's performance in lightweight models.
First, we introduce a network architecture for ASC that utilizes broadcasted residual learning \cite{bcresnet}. Based on this architecture, we can achieve higher accuracy while reducing the size by a third of the baseline \cite{receptivefield}. Next, we propose a novel normalization method, Residual Normalization (ResNorm), \sh{which} can leverage the generalization performance for unseen devices. ResNorm allows maintaining classification accuracy while minimizing the influence on different frequency responses of devices by performing normalization of frequency bands in the residual path. 
Finally, we describe model compression combined with pruning and quantization to satisfy the model complexity of the task while maintaining performance using knowledge distillation.

This work is an expanded version from the challenge technical report submissions \cite{Kim2021b}. The rest of the paper is organized as follows. Section 2 describes the network architecture, Residual Normalization, and model compression methods. Section 3 shows the experimental results and analysis. Finally, we conclude the work in Section 4.

\section{Proposed Method}
\label{sec:method}
This session introduces an efficient model design for 
device-imbalanced acoustic scene classification. First, we \sm{present} a modified version of Broadcasting-residual network \cite{bcresnet} for the acoustic scene domain. Following, we propose Residual Normalization for generalization in a device-imbalanced dataset. Finally, we describe how to get a compressed version of the proposed system.

\subsection{Network Architecture}
\label{sec:architecture}

To design a low-complexity network in terms of the number of parameters, we use \sm{a}  Broadcasting-residual network (BC-ResNet) \cite{bcresnet} which uses 1D and 2D CNN feature\sh{s} together for better efficiency. While the BC-ResNet targets human voice, we \sh{aim to} classify the audio scenes.
\sh{To adapt to the differences in input domains, we make two modifications to the network, {\it i.e.}, limit the receptive field and use max-pool instead of dilation.}

\begin{table}[t]
    \caption{\textbf{BC-ResNet-ASC.} Each row is a sequence of one or more identical modules repeated $n$ times with input shape of frequency by time by channel and total time step $T$.}
    \label{architecture}
    \centering
    \resizebox{\linewidth}{!}{
    \setlength{\tabcolsep}{1em}
    \begin{tabular}{c|c|c|c}
    \toprule
    Input 
    & Operator & n & Channels \\
    \midrule
    $256 \times T \times 1$ & conv2d 5x5, stride 2& - & 2c\\
    $128 \times T/2 \times 2c$ & stage1: BC-ResBlock & 2 & c\\
    $128 \times T/2 \times c$ & max-pool 2x2 & - & - \\
    $64 \times T/4 \times c$ & stage2: BC-ResBlock & 2 & 1.5c\\
    $64 \times T/4 \times 1.5c$ & max-pool 2x2 & - & - \\
    $32 \times T/8 \times 1.5c$ & stage3: BC-ResBlock & 2 & 2c\\
    $32 \times T/8 \times 2c$ & stage4: BC-ResBlock & 3 & 2.5c\\
    $32 \times T/8 \times 2.5c$ & conv2d 1x1 & - & num class\\
    $32 \times T/8 \times$ num class & avgpool & - & - \\
    $1 \times 1 \times$ num class & - & - & - \\
    \bottomrule
    \end{tabular}
    }
\end{table}

\wh{The proposed architecture, BC-ResNet-ASC, is shown in Table~\ref{architecture}.}
The model has 5x5 convolution on the front with a (2, 2) stride for downsampling followed by BC-ResBlocks \cite{bcresnet}. In \cite{receptivefield}, they show that the size of the receptive field can regularize CNN-based ASC models. We change the depth of the network and use max-pool to control the size of the receptive field. With a total of 9 BC-ResBlocks and two max-pool layers, the receptive field size is 109x109. We also do the last 1x1 convolution before global average pooling that the model classifies each receptive field separately and ensembles them by averaging. Original BC-ResNets use dilation in temporal dimension to obtain a larger receptive field while maintaining temporal resolution across the network. We observe that time resolution does not need to be fully kept in the audio scene domain, and instead of dilation, we insert max-pool layers in the middle of the network.

In this work, we use \textit{BC-ResNet-ASC-1}  and \textit{BC-ResNet-ASC-8} whose base number\wh{s} of channels $c$ are 10 and 80, respectively, in Table~\ref{architecture}.
Table~\ref{architecture_score} compares our BC-ResNet-ASC-8 with two baselines: CP-ResNet \cite{receptivefield} which is a residual network-based ASC model with limited receptive field size; and original BC-ResNet-8 with the number of Subspectral Normalization \cite{ssn} groups of 4. As shown in Table~\ref{architecture_score}, BC-ResNet-ASC-8 records Top-1 test accuracy 69.5\% with only one-third number of parameters compared to CP-ResNet showing 67.8\% accuracy. Moreover, BC-ResNet-ASC-8 outperforms the original BC-ResNet-8 by a 1\% margin with the modifications.


\subsection{Residual Normalization}
Instance normalization (IN) \cite{instancenorm} is a representative approach to reducing unnecessary domain gaps for better domain generalization \cite{batchinstancenorm} or domain style transfer \cite{adain, jung2020arbitrary} in the image domain.
While domain difference can be captured by channel mean and variance in \sh{the} image domain, we observe that differences between audio devices are revealed along frequency dimension as shown in Figure~\ref{fig:tsneplot}. To get audio device generalized features, we use instance normalization by frequency (FreqIN) as below.
\begin{equation}
FreqIN(x) = \frac{x-\mu_{nf}}{\sqrt{\sigma_{nf}^{2} + \epsilon}},
\label{eq:freqin}
\end{equation}
where,
\begin{align}
\mu_{nf} &= \frac{1}{CT}\sum_{c=1}^{C}\sum_{t=1}^{T}x_{ncft}, \nonumber \\ 
\sigma^2_{nf} &= \frac{1}{CT}\sum_{c=1}^{C}\sum_{t=1}^{T}{(x_{ncft}-\mu_{nf})^2}. 
\label{eq:meanvar}
\end{align}
\sm{Here, $\mu_{nf}$, $\sigma_{nf} \in \mathbb{R}^{N\times F}$ are mean and standard deviation of the input feature $x \in \mathbb{R}^{N\times C\times F\times T}$, where $N$, $C$, $F$, $T$ denote batch size, number of channel, frequency dimension, and time dimension respectively.}
\sh{$\epsilon$ is a small number added to avoid division by zero.}


\begin{table}[t]
    \caption{\textbf{Network Architectures.} Compare Top-1 test accuracy (\%) on TAU Urban AcousticScenes 2020 Mobile, development dataset.}
    \label{architecture_score}
    \centering
    \resizebox{\linewidth}{!}{
    \setlength{\tabcolsep}{1em}
    \begin{tabular}{l|c|c}
    \toprule
    Network Architecture & \#Param & Top-1 Acc. (\%) \\
    \midrule
    CP-ResNet, c=64 & 899k & 67.8\\
    BC-ResNet-8, $\text{num SSN group}=4$ & 317k & 68.6 $\pm$ 0.4\\
    BC-ResNet-ASC-8 & 315k & 69.5 $\pm$ 0.3\\
    \bottomrule
    \end{tabular}
    }
\end{table}

\begin{figure}[t]
  \centering
  \includegraphics[width=1.\columnwidth]{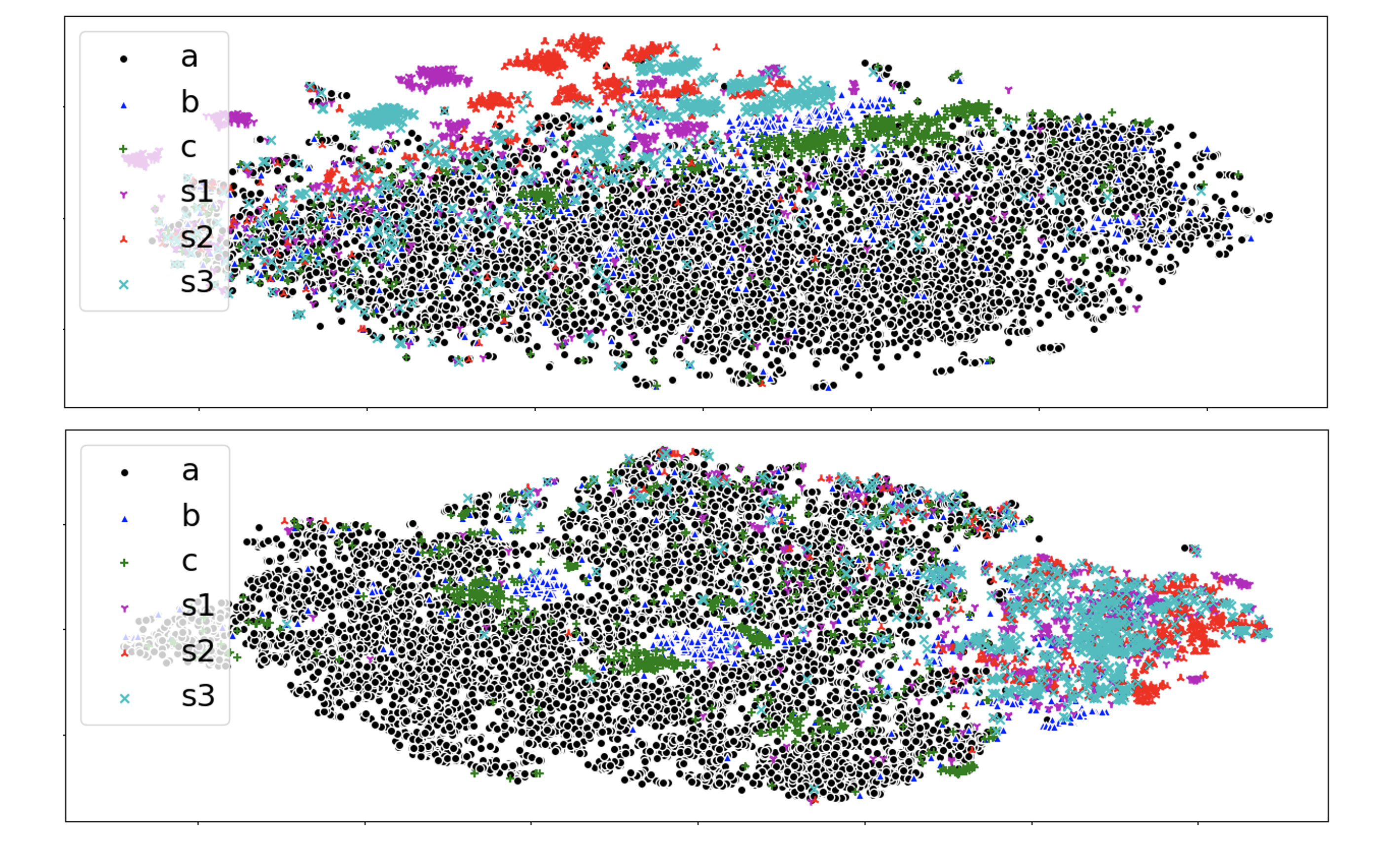}
  \caption{\textbf{2D t-SNE \cite{tsne} visualization} of feature maps of BC-ResNet-ASC-1 stage2 (without ResNorm). \textbf{Top:} Concatenation of frequency-wise mean and standard deviations. \textbf{Bottom:} Concatenations of channel mean and standard deviations. The training samples are separated better by device ID (A to S3) with frequency-wise statistics.
  }
  \label{fig:tsneplot}
\end{figure}

Direct use of IN can result in loss of useful information for classification contained in domain information. To compensate for information loss due to FreqIN, we add an identity shortcut path multiplied by a hyperparameter $\lambda$. We suggest a normalization method, named Residual Normalization (ResNorm) which is
\begin{equation}
\textit{ResNorm}(x) = \lambda \cdot x + \textit{FreqIN}(x).
\label{eq:resnorm}   
\end{equation}
We apply ResNorm for input features and after the end of every stage in Table~\ref{architecture}. There are a total of five ResNorm modules in the network.

\begin{table*}[t]
    \caption{\textbf{Residual Normalization.} We demonstrate how residual normalization affects BC-ResNet-ASC on TAU Urban AcousticScenes 2020 Mobile, development dataset. We show mean and standard deviation of Top-1 test accuracy (\%) (averaged over 3 seeds, * averaged over 6 seeds).}
    \label{result_table}
    \centering
    \resizebox{\linewidth}{!}{
    \begin{tabular}{l|c|ccccccccc|c}
    \toprule
    Method & \#Param & A & B & C & S1 & S2 & S3 & S4 & S5 & S6 & Overall  \\
    \midrule
    BC-ResNet-ASC-1 (Baseline) & 8.1k & 73.1 & 61.2 & 65.3 & 58.2 & 57.3 & 66.2 & 51.5 & 51.5 & 46.3 & 58.9 $\pm$ 0.8 \\
    BC-ResNet-ASC-1 + Global FreqNorm & 8.1k & 73.9 & 60.9 & 65.5 & 60.2 & 57.9 & 67.9 & 50.2 & 54.3 & 49.4 & 60.0 $\pm$ 0.9\\
    BC-ResNet-ASC-1 + Fixed PCEN & 8.1k & 68.0 & 60.4 & 57.2 & 64.0 & 63.0 & 66.2 & 62.3 & 61.8 & 56.5 & 62.2 $\pm$ 0.8\\
    \midrule
    \textbf{BC-ResNet-ASC-1 + ResNorm} & 8.1k & \textbf{76.4} & 65.1 & \textbf{68.3} & \textbf{66.0} & 62.2 & \textbf{69.7} & 63.0 & 63.0 & 58.3 & *65.8 $\pm$ 0.7\\
    ~~ w/o ResNorm in Network & 8.1k & 75.1 & \textbf{68.9} & 67.0 & \textbf{66.0} & 63.9 & 69.3 & 63.4 & \textbf{66.9} & \textbf{63.6} & \textbf{67.1 $\pm$ 0.8}\\
    ~~ w/o Shortcut & 8.1k & 68.2 & 62.1 & 58.6 & 64.2 & \textbf{65.3} & 66.3 & \textbf{65.1} & 63.8 & 61.3 & 63.9 $\pm$ 0.7\\
    \midrule
    \textbf{BC-ResNet-ASC-8 + ResNorm} & 315k & \textbf{81.3} & \textbf{74.4} & \textbf{74.2} & \textbf{75.6} & \textbf{73.1} & \textbf{78.6} & 73.0 & \textbf{74.0} & \textbf{72.7} & \textbf{*75.2 $\pm$ 0.4} \\
    ~~ w/o ResNorm in Network & 315k & 80.8 & 73.7 & 73.0 & 74.0 & 72.9 & 77.8 & \textbf{73.3} & 72.1 & 71.0 & 74.3 $\pm$ 0.3 \\
    ~~ w/o Shortcut & 315k & 78.3 & 73.5 & 69.1 & 73.8 & 72.9 & 75.6 & 72.2 & 72.5 & 71.0 & 73.2 $\pm$ 0.3\\
    \bottomrule
    \end{tabular}
    }
\end{table*}

\subsection{Model Compression}
\label{model_compression}
To compress the proposed model, we utilize three model compression schemes: pruning, quantization, and knowledge distillation.

\noindent \textbf{Pruning.} The pruning method prunes unimportant weights or channels based on many criteria. In this work, we choose a magnitude-based one-shot unstructured pruning scheme used in \cite{NEURIPS2020_eb1e7832}. \sm{After training, we conduct unstructured pruning on all convolution layers and do additional training to enhance the pruned model's performance.}

\noindent \textbf{Quantization.} 
\sm{Quantization is the method to map continuous infinite values to a smaller set of discrete finite values.}
We quantize all of our models with quantization-aware training (QAT) with symmetric quantization \cite{NEURIPS2020_eb1e7832}. We combine the pruning and quantization methods. \wh{It means that} we quantize the important weights which are not pruned after the pruning process in the additional training phase. We quantize all convolution layers as an 8-bit while utilize the half-precision representation for other weights.

\noindent \textbf{Knowledge Distillation.} Knowledge Distillation (KD) trains the lightweight model using \wh{the outputs of} a pre-trained teacher network.
In general, previous model compression schemes such as pruning and quantization decrease the performance by reducing the model complexity.
To enhance the performance of the compressed model, we use a KD loss \cite{kim2021feature} using the pre-trained model as a teacher network. 

\section{Experiments}
\label{sec:experiment}
\subsection{Experimental Setup}
\noindent \textbf{Datasets.} We evaluate the proposed method on the TAU Urban Acoustic Scenes 2020 Mobile, development dataset \cite{dcase_dataset}. The dataset consists of a total of 23,040 audio segment recordings from 12 European cities in 10 different acoustic scenes using 3 real devices (A, B, and C) and 6 simulated devices (S1-S6). The 10 acoustic scenes contain ``airport'', ``shopping mall'', ``metro station'', ``pedestrian street'', ``public square'', ``street with traffic'', ``park'', and travelling by ``tram'', ``bus'', and ``metro''. Audio segments from B and C are recorded simultaneously with device A, but not perfectly synchronized.
Simulated devices S1-S6 generate data using randomly selected audio segments from real device A.
Each utterance is 10-sec-long and the sampling rate is 48kHz.
\cite{dcase_dataset} divides the dataset into training and test of 13,962 and 2,970 segments, respectively. In the training data, device A has 10,215 samples while B, C, and S1-S3 have 750 samples each, which means the data is device\sh{-}imbalanced. Devices S4-S6 remain unseen in training. In test data, all devices from A to S6 have 330 segments each.

\noindent \textbf{Implementation Details.} We do downsampling by 16kHz and use input features of 256-dimensional log Mel spectrograms with a window length of 130ms and a frameshift of 30ms. During training, we augment data to get a more generalized model. In the time dimension, we randomly roll each input feature in the range of -1.5 to 1.5 sec, and the out-of-range part is added to the opposite side. We also use Mixup \cite{mixup} with $\alpha=0.3$ and Specaugment \cite{specaugment} with two frequency masks and two temporal masks with mask parameters of 40 and 80\sm{, respectively, }except time warping. We use Specaugment only for the large model, BC-ResNet-ASC-8. In BC-ResNet-ASC, we use Subspectral Normalization \cite{ssn} as indicated in \cite{bcresnet} with 4 sub-bands and use dropout rate of 0.1. We train the models for 100 epoch using stochastic gradient descent (SGD) optimizer with momentum to 0.9, weight decay to 0.001, mini-batch size to 64, and learning rate linearly increasing from 0 to 0.06 over the first five epochs as \sm{a} warmup \cite{warmup} before decaying to zero with cosine annealing \cite{cosine_schedule} for the rest of the training. We use fixed $\lambda = 0.1$ for ResNorm in experiments. Due to the absence of validation split in the development dataset, we report the numbers of early stopping.

\noindent \textbf{Baselines.} We compare our method with other methods and do some ablation studies: 1) Global FreqNorm, which normalizes data by global mean and variance of each frequency bin;
2) Fixed per-channel energy normalization (PCEN) \cite{pcen}, which is an automatic gain control based dynamic compression and is used instead of log Mel spectrogram in our experiment; 3) \textit{w/o ResNorm in Network}, which uses ResNorm module only at input not in the middle of the network. 4) \textit{w/o shortcut}, which is a special case of ResNorm when $\lambda=0$ in Equation~\ref{eq:resnorm} and uses FreqIN.

\subsection{Residual Normalization}

We do the experiments using BC-ResNet-ASC-1 and BC-ResNet-ASC-8, and the overall results are on Table~\ref{result_table}. The task has multi-device inputs which are imbalanced with dominant device A. As a result, the baseline, BC-ResNet-ASC-1, shows that the accuracy of the device A is relatively higher than other seen devices, B, C, S1, S2, and S3.
\wh{Furthermore, the accuracy on unseen devices, S4, S5, and S6 are even lower, and these results imply that the model is not generalized well to multiple devices, especially for unseen devices.}
When we use global normalization by frequency dimension, the result shows 60.0\% accuracy which is 1\% improvements compared to the baseline, but still we can observe poor domain generalization. We also try PCEN, a normalized feature instead of log Mel spectrogram. PCEN shows improvements for unseen devices, but we also observe that the performance of device A degrades due to its normalization.
The proposed ResNorm use\wh{s} FreqIN to get domain invariant features while not loosing the useful \wh{class-discriminative} information through identity shortcut connection. The `BC-ResNet-ASC-1 + ResNorm' shows a large improvement, 6\% compared to baseline and records 65.8\% test accuracy. The ResNorm shows performance improvements not just for unseen devices but also for all seen devices.

We do some ablation studies for the component of ResNorm. First, we use \wh{the} ResNorm module as \wh{the} preprocessing module, and do not use the module in the middle of the network; `w/o ResNorm in Network'.
For the small model, BC-ResNet-ASC-1, `w/o ResNorm in Network' shows better performance, 67.1\%, and for the larger model, BC-ResNet-ASC-8, it shows a performance degradation of 1\%. Due to ResNorm's regularization effect, it was expected that this module could degrade the performance of a small network. We expect that the module can control the normalization power by the hyperparameter $\lambda$ in Equation~\ref{eq:resnorm} to adapt to various size of networks. In this work, we use fixed $\lambda=0.1$, and leave the automatic update of the $\lambda$ as a future work. Second, `w/o shortcut' shows the result when $\lambda=0$ in ResNorm which equals to FreqIN in Equation~\ref{eq:freqin}. Our design motivation is that the shortcut path will keep the useful information for classification. The results show that FreqIN records relatively lower accuracy for seen devices compared to ResNorm. Especially, the margins on device A are 8.2\% and 3.0\% on BC-ResNet-ASC-1 and BC-ResNet-ASC-8, respectively.

\begin{table}[t]
\centering
  \caption{\textbf{Model compression} Compare bitwidth, top-1 test accuracy (\%) on Tau Urban AcousticScenes 2020 Mobile, development dataset, and pruning ratio of the models (Average over 6 seeds).}
\label{table_model_compression}
\begin{adjustbox}{width=1\linewidth}
  \begin{tabular}{l c c c c}
        \toprule
        \multicolumn{5}{c}{BC-ResNet-ASC-8 + ResNorm, 300 epochs, KD} \\
        \midrule
            Method & Bitwidth & KD & Pruning & Accuracy  \\
        \midrule
            Vanilla model & 32 & - & - & 76.3 $\pm$ 0.8 \\
        \midrule
            Compressed model & 8, 16 & & 0.89 & 75.1 $\pm$ 0.9 \\
            Compressed model & 8, 16 & \checkmark  & 0.89 & 75.3 $\pm$ 0.8 \\
        \bottomrule
  \end{tabular}
\end{adjustbox}
\end{table}

\subsection{Model Compression}
\wh{Simultaneously, we distill the knowledge of the pre-trained teacher network (`Vanilla' model) into the compressed model for enhancing the performance and achieve the 0.2\% improvement in test accuracy.
In detail,} we prune the convolution layers of the model with 89\% pruning ratios compared to vanilla \sm{and} quantize all convolution layers in a compressed model as an 8-bit. Other layers are quantized as a 16-bit. The resulting `Compressed' model has 33K 8-bit nonzero for convolution layers and 15K 16-bit parameters for normalization, resulting in 61.5kB and shows 75.3\% test accuracy which is 1\% lower than Vanilla model. We use the ensemble of two compressed model in the DCASE 2021 challenge, task 1A.  

\section{Conclusions}
\label{sec:conclude}

In this work, we design a system to achieve two goals; 1) efficient design in terms of the number of parameters and 2) adapting to device imbalanced dataset. To design an efficient acoustic scene classification model, we suggest a modified version of Broadcasting residual network \cite{bcresnet} by limiting receptive field and using max-pool. We compress the model further by utilizing three model compression schemes, pruning, quantization, and knowledge distillation. Moreover, we propose a frequency-wise normalization method, named Residual Normalization which uses instance normalization by frequency and shortcut connection to be generalized to \sm{multiple devices} while not losing \sm{discriminative} information. Our system achieves 76.3\%
test accuracy on TAU Urban \sm{Acoustic Scenes 2020 Mobile}, development dataset with 315k number of parameters and the compressed version \sm{achieves} 75.3 \% test accuracy with 89\% pruning, 8-bit quantization, and knowledge distillation. 
Residual normalization has a hyperparameter $\lambda$ which can control the regularization power of the module. We leave the automatic update of the hyperparameter as future work.

\bibliographystyle{IEEEtran}
\bibliography{refs}

\begin{thebibliography}{10}
\providecommand{\url}[1]{#1}
\def\UrlFont{\rmfamily}
\providecommand{\newblock}{\relax}
\providecommand{\bibinfo}[2]{#2}
\providecommand\BIBentrySTDinterwordspacing{\spaceskip=0pt\relax}
\providecommand\BIBentryALTinterwordstretchfactor{4}
\providecommand\BIBentryALTinterwordspacing{\spaceskip=\fontdimen2\font plus
\BIBentryALTinterwordstretchfactor\fontdimen3\font minus
  \fontdimen4\font\relax}
\providecommand\BIBforeignlanguage[2]{{%
\expandafter\ifx\csname l@#1\endcsname\relax
\typeout{** WARNING: IEEEtran.bst: No hyphenation pattern has been}%
\typeout{** loaded for the language `#1'. Using the pattern for}%
\typeout{** the default language instead.}%
\else
\language=\csname l@#1\endcsname
\fi
#2}}

\bibitem{valenti2016dcase}
M.~Valenti, A.~Diment, G.~Parascandolo, S.~Squartini, and T.~Virtanen,
  ``{DCASE} 2016 acoustic scene classification using convolutional neural
  networks,'' in \emph{Proc. Workshop Detection Classif. Acoust. Scenes
  Events}, 2016, pp. 95--99.

\bibitem{radhakrishnan2005audio}
R.~Radhakrishnan, A.~Divakaran, and A.~Smaragdis, ``Audio analysis for
  surveillance applications,'' in \emph{IEEE Workshop on Applications of Signal
  Processing to Audio and Acoustics, 2005.}\hskip 1em plus 0.5em minus
  0.4em\relax IEEE, 2005, pp. 158--161.

\bibitem{chu2009environmental}
S.~Chu, S.~Narayanan, and C.-C.~J. Kuo, ``Environmental sound recognition with
  time--frequency audio features,'' \emph{IEEE Transactions on Audio, Speech,
  and Language Processing}, vol.~17, no.~6, pp. 1142--1158, 2009.

\bibitem{dcase2021web}
\url{http://dcase.community/challenge2021/}.

\bibitem{dcase_task1A}
I.~Martín-Morató, T.~Heittola, A.~Mesaros, and T.~Virtanen, ``Low-complexity
  acoustic scene classification for multi-device audio: analysis of dcase 2021
  challenge systems,'' 2021.

\bibitem{dcase_dataset}
\BIBentryALTinterwordspacing
T.~Heittola, A.~Mesaros, and T.~Virtanen, ``Acoustic scene classification in
  dcase 2020 challenge: generalization across devices and low complexity
  solutions,'' in \emph{Proceedings of the Detection and Classification of
  Acoustic Scenes and Events 2020 Workshop (DCASE2020)}, 2020, submitted.
  [Online]. Available: \url{https://arxiv.org/abs/2005.14623}
\BIBentrySTDinterwordspacing

\bibitem{task1a2020best_cnn}
S.~Suh, S.~Park, Y.~Jeong, and T.~Lee, ``Designing acoustic scene
  classification models with {CNN} variants,'' DCASE2020 Challenge, Tech. Rep.,
  June 2020.

\bibitem{receptivefield}
K.~Koutini, H.~Eghbal{-}zadeh, M.~Dorfer, and G.~Widmer, ``The receptive field
  as a regularizer in deep convolutional neural networks for acoustic scene
  classification,'' in \emph{{EUSIPCO}}.\hskip 1em plus 0.5em minus 0.4em\relax
  {IEEE}, 2019, pp. 1--5.

\bibitem{task1a2020_2nd_cnn}
H.~Hu, C.-H.~H. Yang, X.~Xia, X.~Bai, X.~Tang, Y.~Wang, S.~Niu, L.~Chai, J.~Li,
  H.~Zhu, F.~Bao, Y.~Zhao, S.~M. Siniscalchi, Y.~Wang, J.~Du, and C.-H. Lee,
  ``Device-robust acoustic scene classification based on two-stage
  categorization and data augmentation,'' DCASE2020 Challenge, Tech. Rep., June
  2020.

\bibitem{task1a2019best}
H.~Chen, Z.~Liu, Z.~Liu, P.~Zhang, and Y.~Yan, ``Integrating the data
  augmentation scheme with various classifiers for acoustic scene modeling,''
  DCASE2019 Challenge, Tech. Rep., June 2019.

\bibitem{phaye2019subspectralnet}
S.~S.~R. Phaye, E.~Benetos, and Y.~Wang, ``Subspectralnet--using
  sub-spectrogram based convolutional neural networks for acoustic scene
  classification,'' in \emph{ICASSP 2019-2019 IEEE International Conference on
  Acoustics, Speech and Signal Processing (ICASSP)}.\hskip 1em plus 0.5em minus
  0.4em\relax IEEE, 2019, pp. 825--829.

\bibitem{bcresnet}
B.~Kim, S.~Chang, J.~Lee, and D.~Sung, ``{Broadcasted Residual Learning for
  Efficient Keyword Spotting},'' in \emph{Proc. Interspeech 2021}, 2021, pp.
  4538--4542.

\bibitem{Kim2021b}
B.~Kim, S.~Yang, J.~Kim, and S.~Chang, ``{QTI} submission to {DCASE} 2021:
  Residual normalization for device-imbalanced acoustic scene classification
  with efficient design,'' DCASE2021 Challenge, Tech. Rep., June 2021.

\bibitem{ssn}
S.~Chang, H.~Park, J.~Cho, H.~Park, S.~Yun, and K.~Hwang, ``Subspectral
  normalization for neural audio data processing,'' in \emph{ICASSP 2021-2021
  IEEE International Conference on Acoustics, Speech and Signal Processing
  (ICASSP)}.\hskip 1em plus 0.5em minus 0.4em\relax IEEE, 2021, pp. 850--854.

\bibitem{instancenorm}
D.~Ulyanov, A.~Vedaldi, and V.~S. Lempitsky, ``Instance normalization: The
  missing ingredient for fast stylization,'' \emph{CoRR}, vol. abs/1607.08022,
  2016.

\bibitem{batchinstancenorm}
H.~Nam and H.~Kim, ``Batch-instance normalization for adaptively
  style-invariant neural networks,'' in \emph{NeurIPS}, 2018, pp. 2563--2572.

\bibitem{adain}
X.~Huang and S.~J. Belongie, ``Arbitrary style transfer in real-time with
  adaptive instance normalization,'' in \emph{{ICCV}}.\hskip 1em plus 0.5em
  minus 0.4em\relax {IEEE} Computer Society, 2017, pp. 1510--1519.

\bibitem{jung2020arbitrary}
D.~Jung, S.~Yang, J.~Choi, and C.~Kim, ``Arbitrary style transfer using graph
  instance normalization,'' in \emph{2020 IEEE International Conference on
  Image Processing (ICIP)}.\hskip 1em plus 0.5em minus 0.4em\relax IEEE, 2020,
  pp. 1596--1600.

\bibitem{tsne}
L.~Van~der Maaten and G.~Hinton, ``Visualizing data using t-sne.''
  \emph{Journal of machine learning research}, vol.~9, no.~11, 2008.

\bibitem{NEURIPS2020_eb1e7832}
J.~Kim, K.~Yoo, and N.~Kwak, ``Position-based scaled gradient for model
  quantization and pruning,'' in \emph{Advances in Neural Information
  Processing Systems}, vol.~33, 2020, pp. 20\,415--20\,426.

\bibitem{kim2021feature}
J.~Kim, M.~Hyun, I.~Chung, and N.~Kwak, ``Feature fusion for online mutual
  knowledge distillation,'' in \emph{2020 25th International Conference on
  Pattern Recognition (ICPR)}.\hskip 1em plus 0.5em minus 0.4em\relax IEEE,
  2021, pp. 4619--4625.

\bibitem{mixup}
H.~Zhang, M.~Ciss{\'{e}}, Y.~N. Dauphin, and D.~Lopez{-}Paz, ``mixup: Beyond
  empirical risk minimization,'' in \emph{{ICLR} (Poster)}.\hskip 1em plus
  0.5em minus 0.4em\relax OpenReview.net, 2018.

\bibitem{specaugment}
D.~S. Park, W.~Chan, Y.~Zhang, C.~Chiu, B.~Zoph, E.~D. Cubuk, and Q.~V. Le,
  ``Specaugment: {A} simple data augmentation method for automatic speech
  recognition,'' in \emph{{INTERSPEECH}}.\hskip 1em plus 0.5em minus
  0.4em\relax {ISCA}, 2019, pp. 2613--2617.

\bibitem{warmup}
P.~Goyal, P.~Doll{\'a}r, R.~Girshick, P.~Noordhuis, L.~Wesolowski, A.~Kyrola,
  A.~Tulloch, Y.~Jia, and K.~He, ``Accurate, large minibatch sgd: Training
  imagenet in 1 hour,'' \emph{arXiv preprint arXiv:1706.02677}, 2017.

\bibitem{cosine_schedule}
I.~Loshchilov and F.~Hutter, ``{SGDR:} stochastic gradient descent with warm
  restarts,'' in \emph{{ICLR} (Poster)}.\hskip 1em plus 0.5em minus 0.4em\relax
  OpenReview.net, 2017.

\bibitem{pcen}
Y.~Wang, P.~Getreuer, T.~Hughes, R.~F. Lyon, and R.~A. Saurous, ``Trainable
  frontend for robust and far-field keyword spotting,'' in
  \emph{{ICASSP}}.\hskip 1em plus 0.5em minus 0.4em\relax {IEEE}, 2017, pp.
  5670--5674.

\end{thebibliography}

%
%
%
%
%
%
%
%
%

\end{sloppy}
\end{document}